\documentclass{PoS}

\title{Studies of mesic atoms and nuclei}

\ShortTitle{Mesic atoms and nuclei}

\author{Eliahu Friedman, \speaker{Avraham Gal}\\ 
Racah Institute of Physics, The Hebrew University, 91904 Jerusalem, Israel\\ 
E-mail: \email{avragal@savion.huji.ac.il}}

\author{Ale\v{s} Ciepl\'{y}, Jaroslava Hrt\'{a}nkov\'{a}, Ji\v{r}\'{\i} 
Mare\v{s}\\ 
Nuclear Physics Institute, 250 69 Rez, Czech Republic\\} 

\abstract{$K^-$ mesons offer a unique setting where mesic atoms have been 
studied both experimentally and theoretically, thereby placing constraints 
on the possible existence and properties of meson-nuclear quasibound states. 
Here we review progress in this field made recently by the Jerusalem--Prague 
Collaboration using near-threshold $K^-N$ scattering amplitudes generated 
in several meson--baryon coupled channels models inspired by a chiral EFT 
approach. Our own procedure of handling subthreshold kinematics self 
consistently is used to transform these free-space energy dependent amplitudes 
to in-medium density dependent amplitudes from which $K^-$ optical potentials 
are derived. To fit the world data of kaonic atoms, these single-nucleon 
optical potentials are augmented by multi-nucleon terms. It is found that 
only two of the studied models reproduce also the single-nucleon absorption 
fractions available from old bubble chamber experiments. These two models are 
then checked for possible $K^-$ nuclear quasibound states, despite realizing 
that $K^-$ optical potentials are not constrained by kaonic atom data at 
densities exceeding half nuclear-matter density. We find that when such states 
exist, their widths are invariably above 100 MeV, forbiddingly large to allow 
observation. Multi-nucleon absorption is found to be substantial in this 
respect. This suggests that observable strongly bound $K^-$ mesons are limited 
to the very light systems, such as $K^-pp$.}     

\FullConference{XVII International Conference on Hadron Spectroscopy and 
Structure\\ 25-29 September, 2017\\ University of Salamanca, Salamanca, Spain} 

\begin{document}

\section{Introduction and methodology} 
\label{sec:intro} 

\begin{figure}[htb]
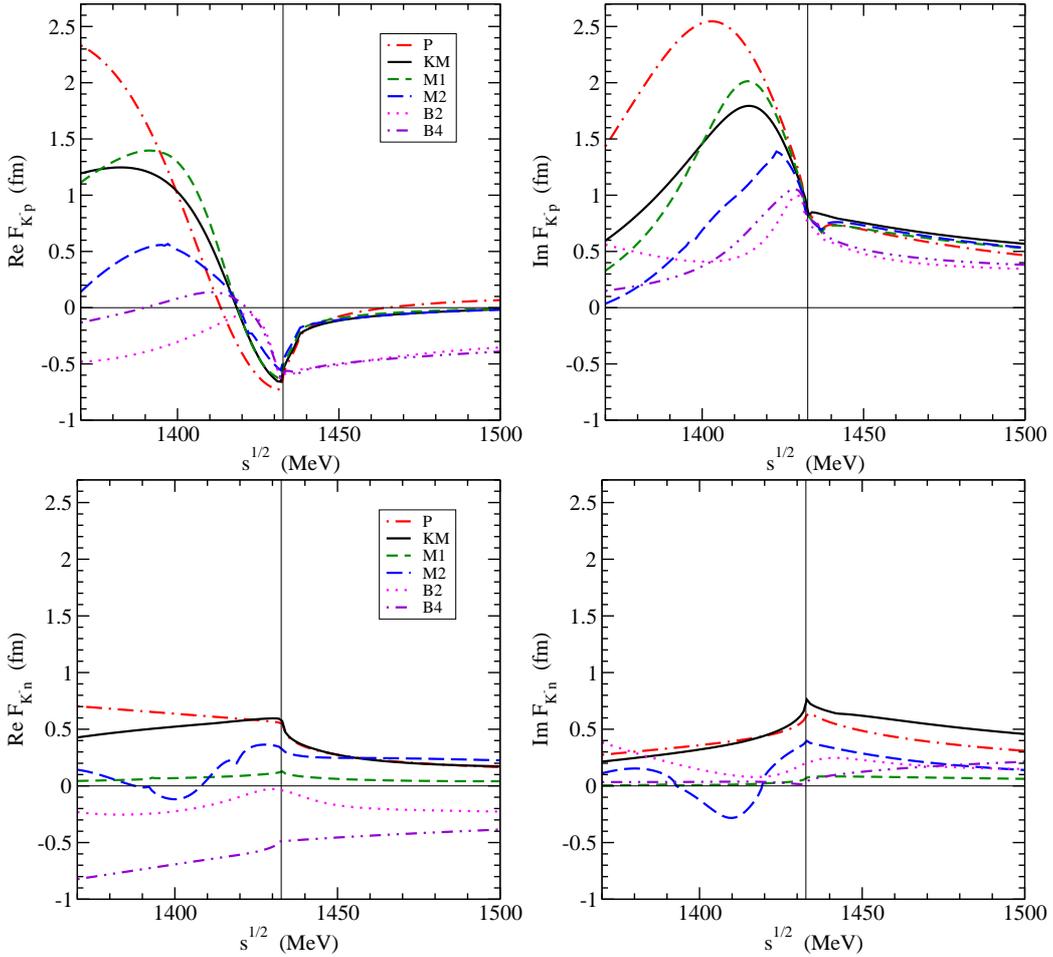
 
\begin{center} 
\includegraphics[width=0.45\textwidth]{ReFkpFree.eps} 
\includegraphics[width=0.45\textwidth]{ImFkpFree.eps} 
\includegraphics[width=0.45\textwidth]{ReFknFree.eps} 
\includegraphics[width=0.45\textwidth]{ImFknFree.eps} 
\caption{Energy dependence of real (left) and imaginary (right) parts 
of $K^-p$ (top) and $K^-n$ (bottom) scattering amplitudes from six 
meson-baryon coupled-channel chirally inspired EFT models~\cite{CMMS16} 
constrained by threshold and low-energy $K^- N$ data. Threshold energies 
$E_{\rm th}$ are marked by vertical lines.} 
\label{fig:Ffree} 
\end{center} 
\end{figure}

Searches for meson ($\overline {K},\eta,\eta^\prime,\omega,\phi$)-nuclear 
quasibound states were reviewed in Hadron2017 by Nanova~\cite{MNP17}. 
Here we focus on $K^-$ mesons for which atomic data provide exclusively 
useful constraints beyond those available from near-threshold $K^-N$ 
data~\cite{BFG97,FG07}. To consider $K^-$ atoms and quasibound nuclear 
states we construct $K^-$ optical potentials based on free-space $K^-N$ 
amplitudes taken from Ref.~\cite{CMMS16} as shown in Fig.~\ref{fig:Ffree}. 
These six amplitude sets exhibit considerable model dependence, particularly 
below threshold, $E<E_{\rm th}$, and for $K^-n$ also above. We note that 
near-threshold meson optical potentials require {\it subthreshold} 
meson-nucleon amplitudes for input~\cite{CFGGM11a,CFGGM11b}. 

\begin{figure}[htb]
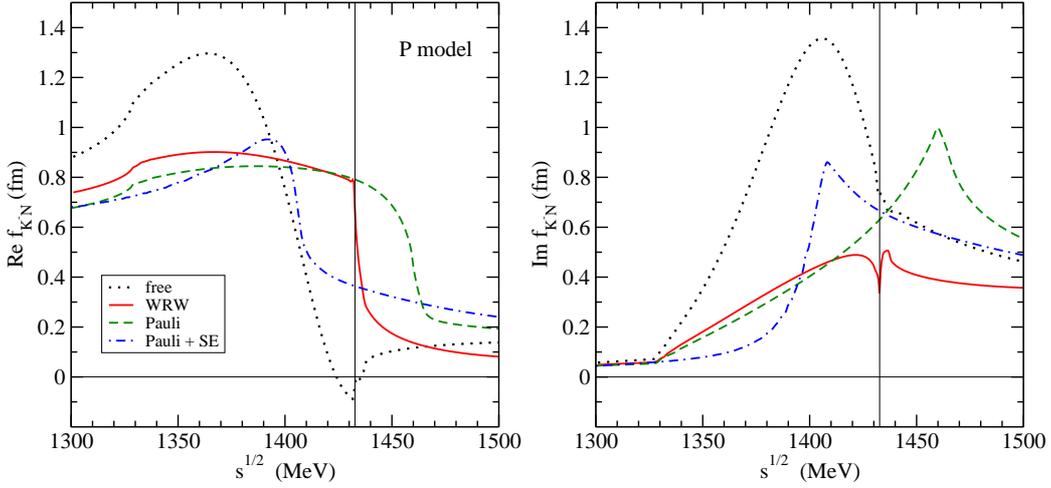

\begin{center}
\includegraphics[width=0.45\textwidth]{ReKN.eps}
\includegraphics[width=0.45\textwidth]{ImKN.eps}
\caption{Energy dependence of the free-space (dotted) P-model amplitudes 
$f_{K^-N}=\frac{1}{2}(f_{K^-p}+f_{K^-n})$, where $f(E)=F(E;\,p=p'=0)$, 
and of several versions of in-medium P-model amplitudes at nuclear-matter 
density $\rho_0=0.17$~fm$^{-3}$ (left: real parts, right: imaginary parts). 
Figure adapted from Ref.~\cite{HM17b}.}
\label{fig:P}
\end{center}
\end{figure}

The free-space amplitudes of Fig.~\ref{fig:Ffree} are then modified to account 
for {\it in-medium} effects, the leading one being the Pauli principle. 
%The energy dependence of several Pauli corrected versions is compared 
The energy dependence of several versions of in-medium amplitudes is compared 
in Fig.~\ref{fig:P} with that of the underlying free-space Prague (P) model 
amplitude from Fig.~\ref{fig:Ffree}. The version applied in our recent 
calculations~\cite{FG17,HM17a,HM17b} is that denoted WRW~\cite{WRW97}. 

\begin{figure}[!h] 
\begin{center} 
\includegraphics[width=0.45\textwidth]{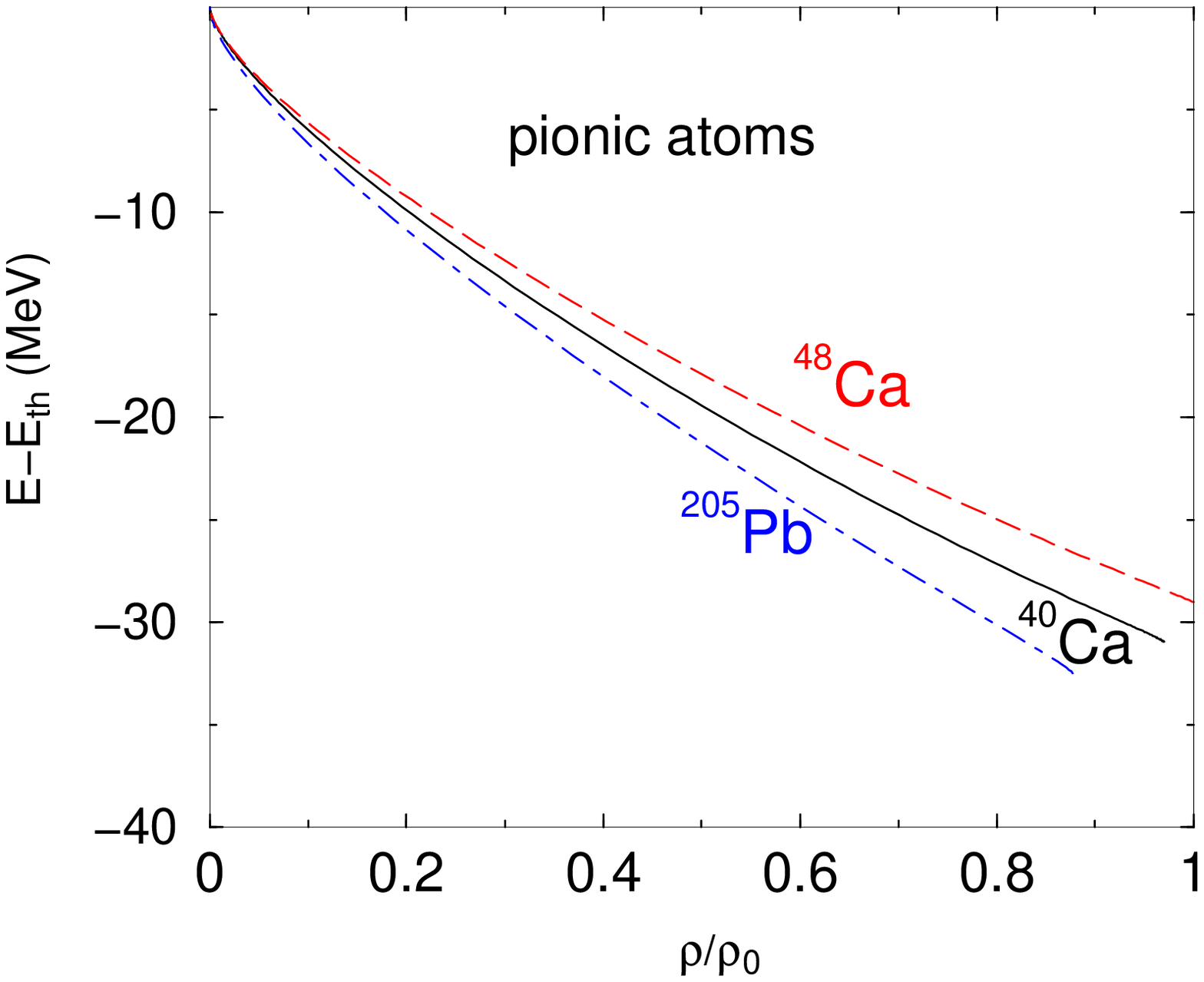} 
\includegraphics[width=0.45\textwidth]{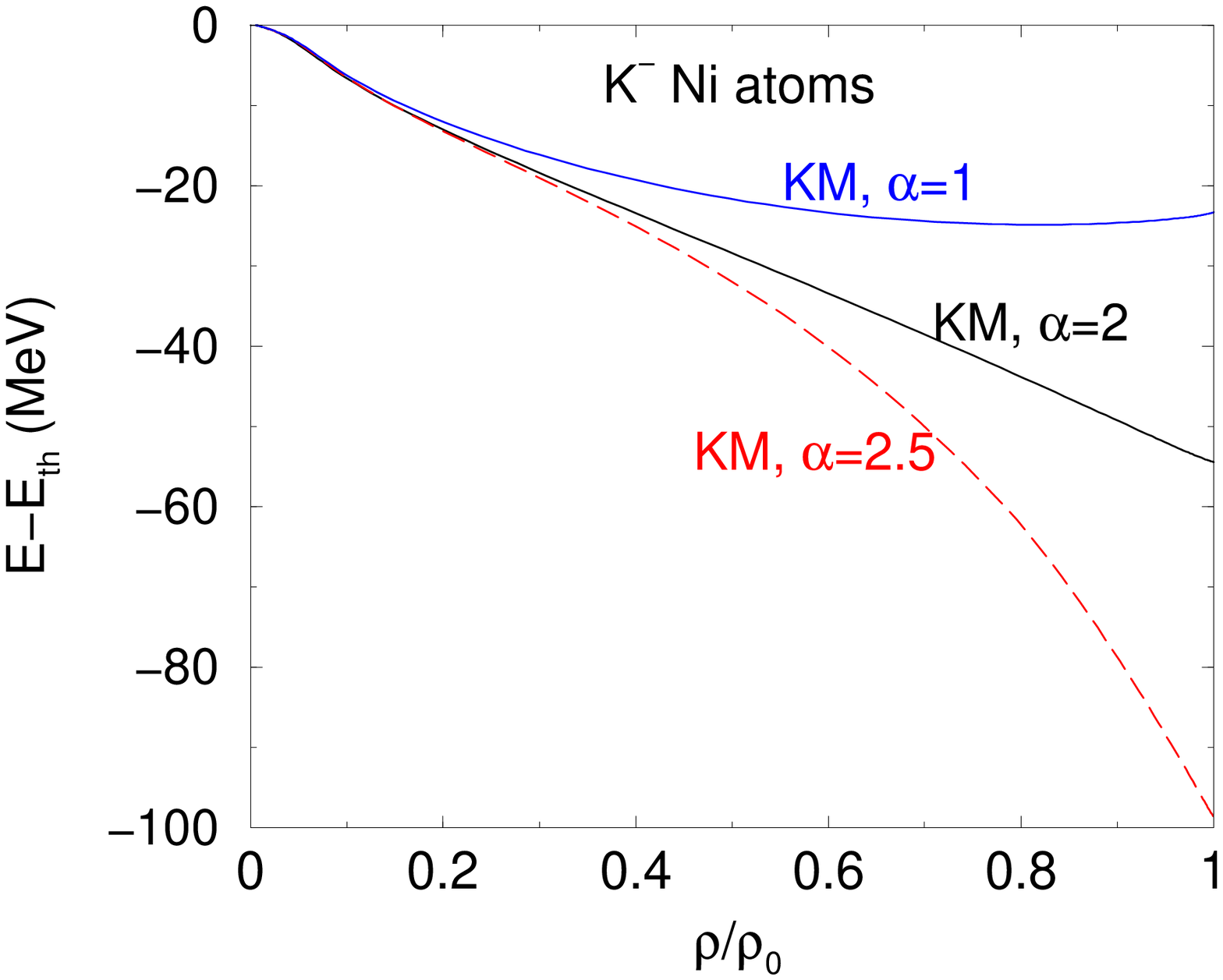} 
\caption{Density dependence of meson-nucleon energy shifts involved in mesic 
atom calculations. The various KM$\alpha$ branches of model KM in the right 
panel are defined in Sect.~\ref{sec:atoms}. Figure adapted from 
Ref.~\cite{FG14}.} 
\label{fig:delsrho} 
\end{center} 
\end{figure}

The subthreshold energy ${\sqrt s}=E_{\rm th}+\delta{\sqrt s}\,$ at which 
hadron optical potentials $V_h(\sqrt s)$ are to be evaluated is then 
determined by solving a self-consistency 
equation~\cite{CFGGM11a,CFGGM11b,FG17,HM17a,HM17b} 
\begin{equation} 
{\delta\sqrt s}(\rho) = -B_N\rho/{\bar\rho} - \beta_N 
[T_N(\rho/\bar{\rho})^{2/3} + B_h\rho/\rho_0 + V_C (\rho/\rho_0)^{1/3}] + 
\beta_h{\rm Re}\,V_h(\sqrt s), 
\label{eq:dels} 
\end{equation} 
in which $V_h(\sqrt s)$=${\tilde t}_{hN}(\sqrt s)\,\rho$ serves as input 
with ${\tilde t}_{hN}$ denoting an in-medium $hN$ $t$-matrix. Next, $V_C$ 
is the $h$-nucleus Coulomb potential, $\beta_{N(h)}=m_{N(h)}/(m_N+m_h)$, 
$B_N=8.5$~MeV, $T_N=23$~MeV and $B_h$ is the hadron binding energy 
obtained by solving the Klein-Gordon equation with potential $V_h(\sqrt s)$. 
The output ${\tilde t}_{hN}$ and $V_h$ have thus become density dependent. 
Fig.~\ref{fig:delsrho} demonstrates a steady decrease of $\delta{\sqrt s}\,
(\rho)$, denoted $E-E_{\rm th}$, for pionic~\cite{FG14} and kaonic~\cite{FG17} 
atoms where $B_h\approx 0$, reaching comparable values of $\delta{\sqrt s}\,$ 
between $-$20 to $-$30~MeV at the nuclear surface ($\rho\approx \frac{1}{2}\, 
\rho_0$).

\section{News from kaonic atoms} 
\label{sec:atoms} 

\begin{figure}[!h] 
\begin{center} 
\includegraphics[width=0.45\textwidth]{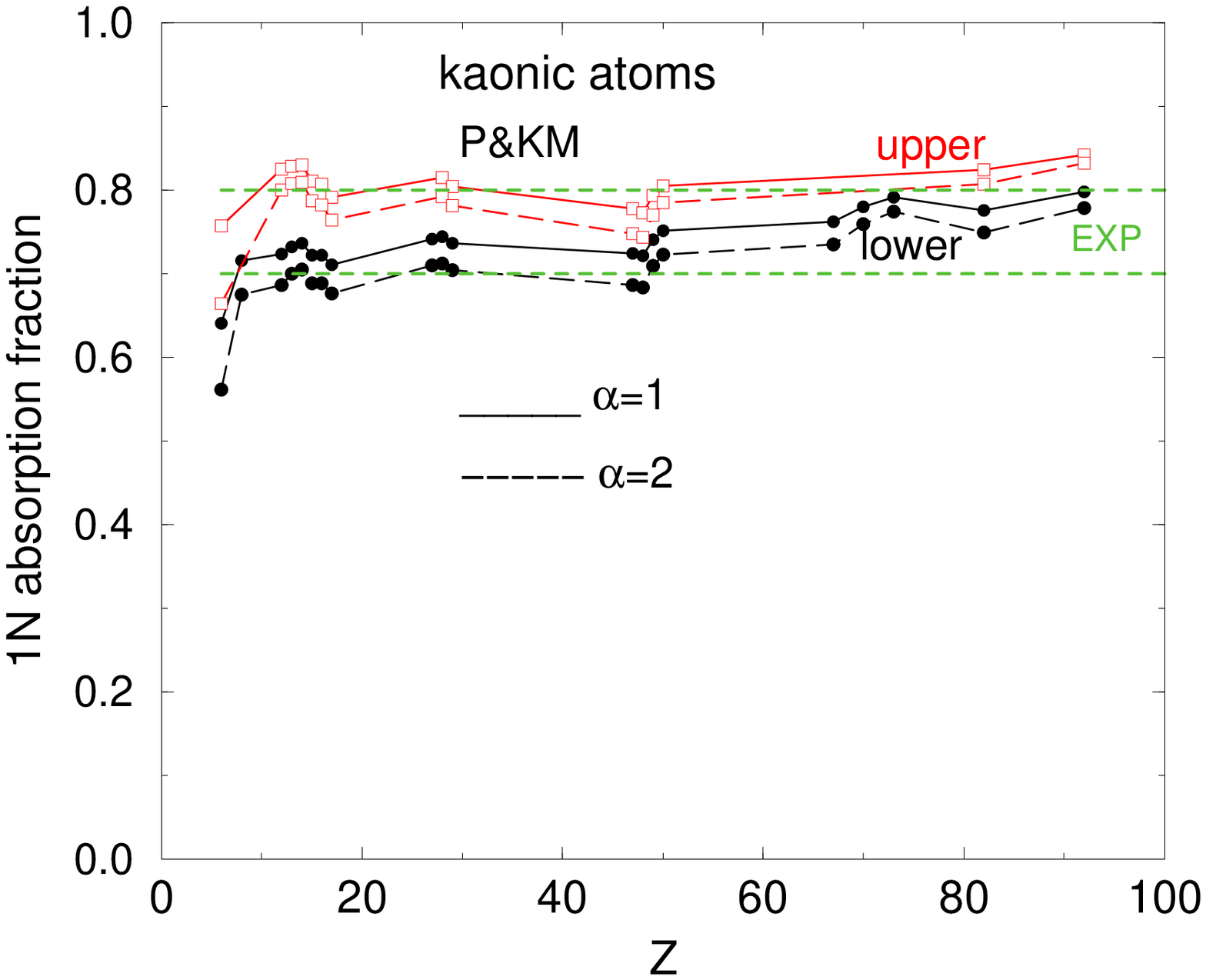} 
\includegraphics[width=0.45\textwidth]{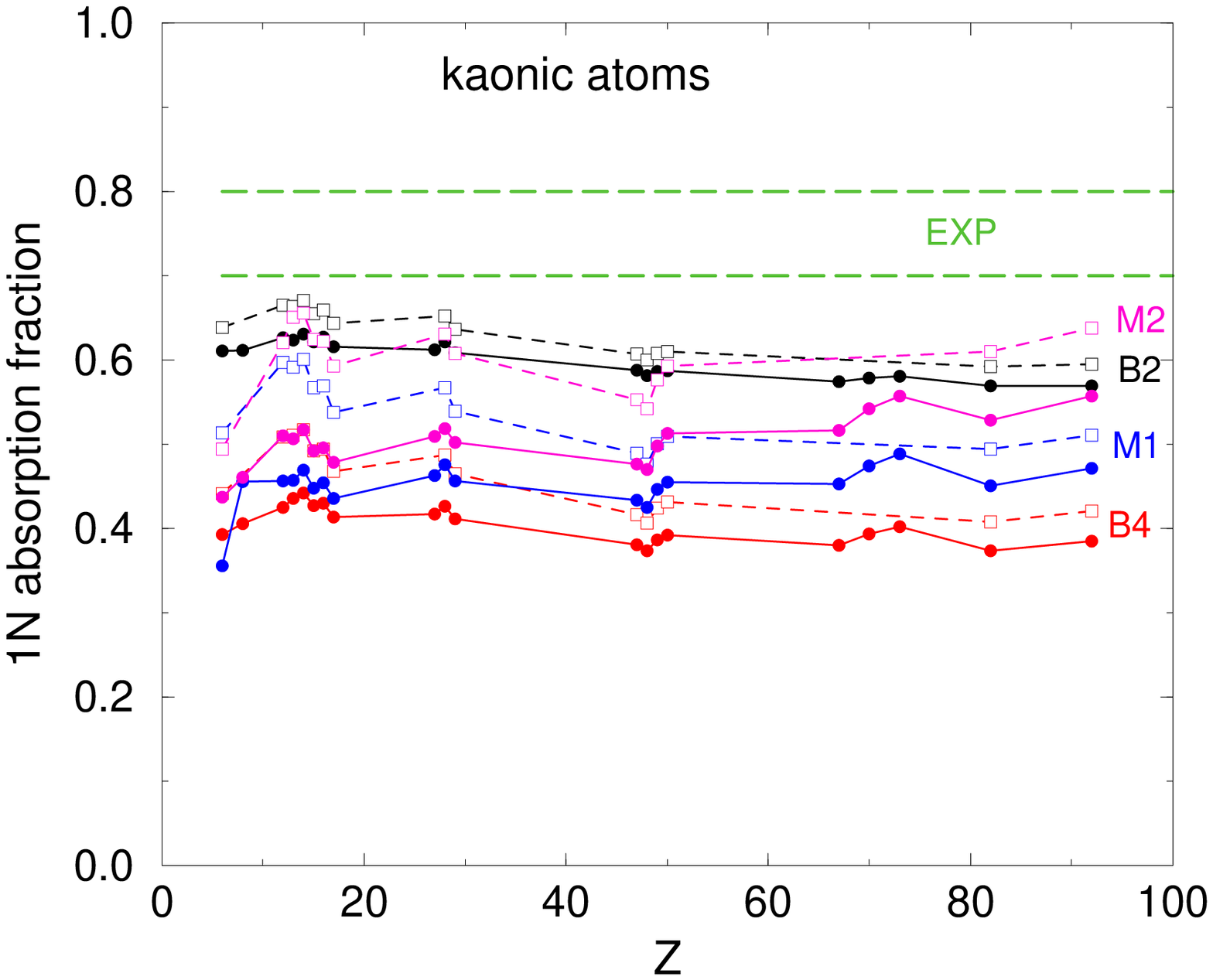} 
\caption{Calculated SNAF \cite{FG17} in models P \& KM (left) and in other 
models (right). For P \& KM, solid circles (open squares) stand for `lower' 
(`upper') states. For right-panel notations and choice of $\alpha$, 
see~\cite{FG17}.} 
\label{fig:SNAF} 
\end{center} 
\end{figure} 

The single-nucleon ($1N$) optical potentials $V_{K^-}^{1N}$=${\tilde t}_{hN}
(\rho)\rho$ of Sect.~\ref{sec:intro}, with in-medium $t$-matrices 
${\tilde t}_{hN}(\rho)$ constructed from free-space $t$-matrices proportional 
to the amplitudes $F$ of Fig.~\ref{fig:Ffree}, fail miserably to fit kaonic 
atom data. However, adding a phenomenological amplitude $B\,(\rho/\rho_0)^{
\alpha}$ to simulate multi-nucleon ($mN$) processes and varying the 
complex strength parameter $B$ upon gridding on the exponent $\alpha$, good 
fits are obtained in all six cases, comparable in quality to published 
phenomenological best fits~\cite{MFG06}. Calculations are then made for 
single-nucleon absorption fractions (SNAF) measured long ago in bubble-chamber 
experiments on nuclear species from C to Br (see~\cite{FG17} for references) 
which are consistent with a common value SNAF$_{\rm exp}\approx 0.75\pm 0.05$. 
These absorption fractions are computed for any atomic state $\psi_j$ by 
splitting $V_{K^-}$ to its $1N$ and $mN$ components, $V_{K^-}=V_{K^-}^{1N}+
V_{K^-}^{mN}$, and evaluating $\Gamma^j_{1N}/\Gamma^j_{\rm tot}$ according to  
\begin{equation} 
\Gamma^j_{\rm tot}=\Gamma^j_{1N}+\Gamma^j_{mN}: \,\,\,\,\,\, 
\Gamma^j_{1N}\sim \int |r\psi_j|^2\,{\rm Im}\,V_{K^-}^{1N}\,dr, \,\,\,\,\,\, 
\Gamma^j_{mN}\sim \int |r\psi_j|^2\,{\rm Im}\,V_{K^-}^{mN}\,dr. 
\label{eq:Im} 
\end{equation} 
Inspecting the overlap of $|r\psi_j|^2$ with Im$\,V_{K^-}$ in the expression 
for $\Gamma^j_{\rm tot}$ one finds that it generally peaks at 15-20\% of 
central nuclear density $\rho_0$ for `lower' states and 10-15\% of $\rho_0$ 
for `upper' states. Fig.~\ref{fig:SNAF} shows very good agreement between 
calculation and experiment for optical potentials based on the $1N$ input of 
models P and KM, including a $mN$ term, and substantial disagreement for the 
other four models. 

\begin{figure}[htb] 
\begin{center} 
\includegraphics[width=0.45\textwidth]{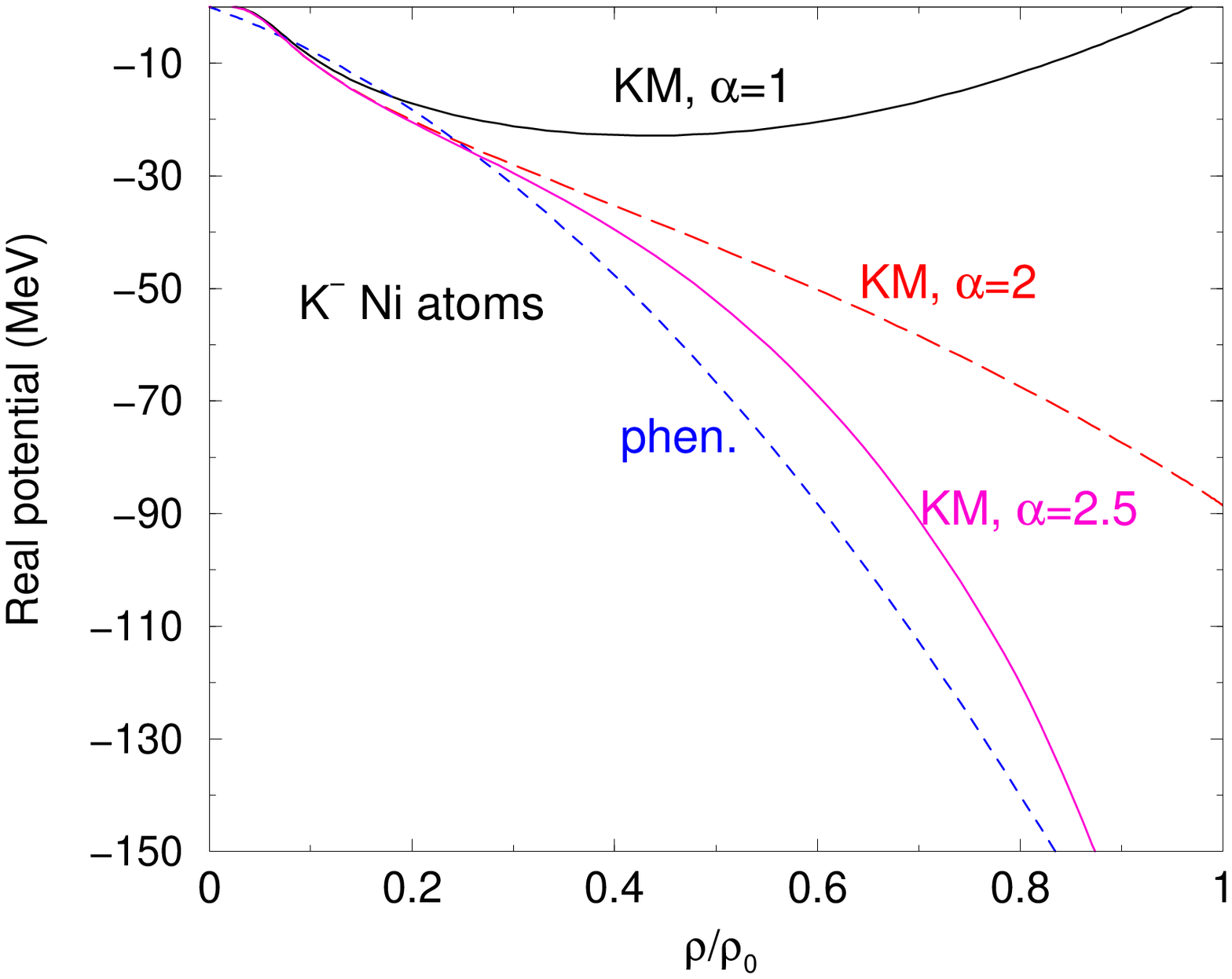} 
\includegraphics[width=0.45\textwidth]{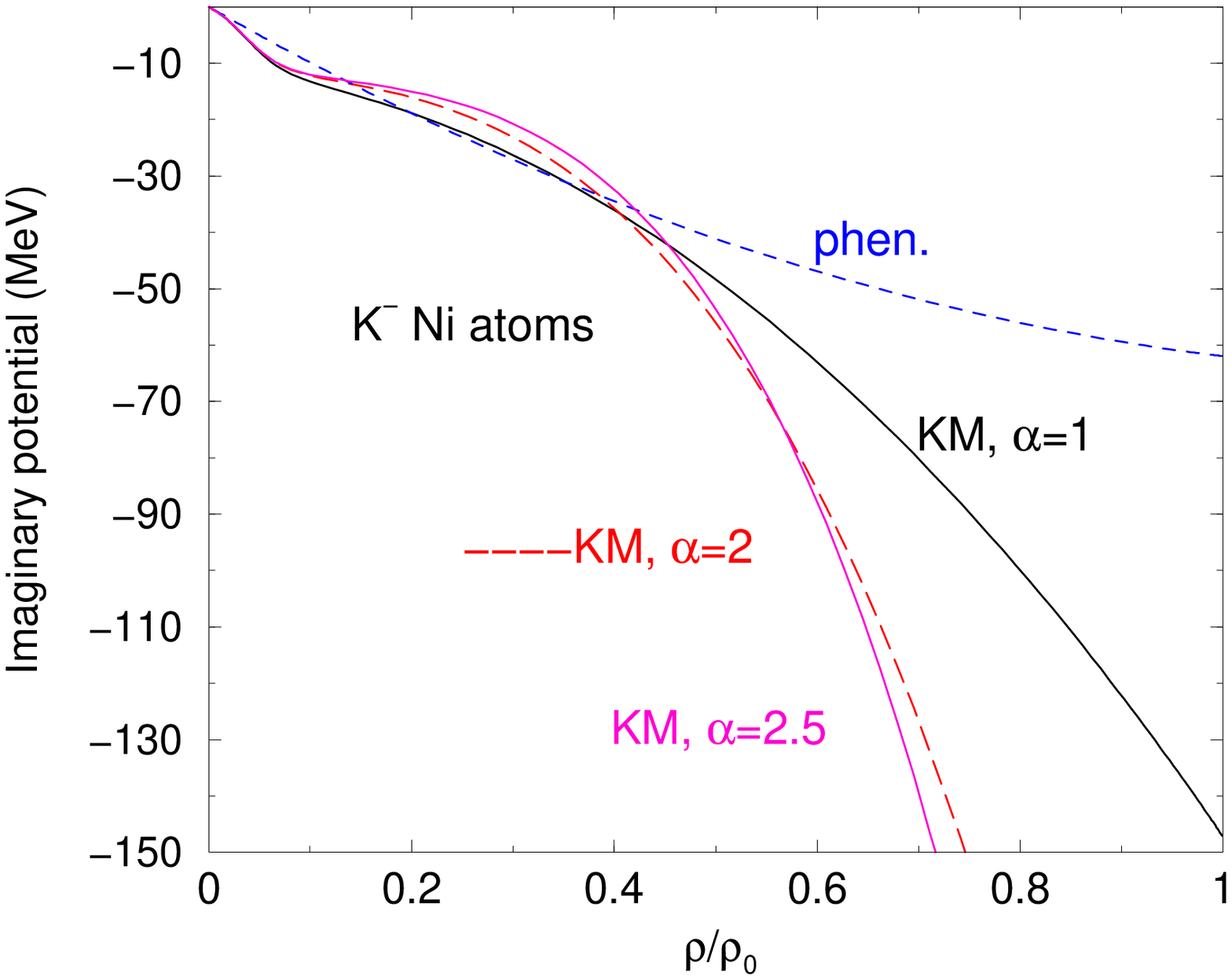} 
\caption{Real part (left) and imaginary part (right) of best-fit $K^-$Ni 
optical potentials `KM$\alpha$' based on the KM $1N$ amplitude plus 
a phenomenological $mN$ amplitude $B\,(\rho/\rho_0)^\alpha$. Shown for 
comparison in short-dashed lines is a {\it purely phenomenological} potential. 
Figure adapted from Ref.~\cite{FG17}.} 
\label{fig:Vopt} 
\end{center} 
\end{figure} 

Fig.~\ref{fig:Vopt} shows $K^-$ optical potentials `KM$\alpha$' in Ni 
atoms, based on the Kyoto-Munich $K^- N$ chiral model amplitude (KM in 
Fig.~\ref{fig:Ffree}) augmented by a phenomenological amplitude $B\,(\rho/
\rho_0)^\alpha$ with three values of exponent $\alpha$, all providing good 
global fits to kaonic atoms data and satisfying the SNAF experimental 
constraint marked by horizontal dashed lines in Fig.~\ref{fig:SNAF}. 
Also shown is a purely phenomenological optical potential producing 
a similarly good fit to kaonic atom data. The real parts of the four 
plotted potentials agree with each other up to about 25\% of the central 
nuclear density. In contrast, the imaginary part is seen to be well determined 
up to 50\% of the central density, reflecting the observation that strong 
interaction effects in kaonic atoms, where level widths are significantly 
larger than shifts, are dominated by the imaginary potential. Hence we 
conclude that at larger densities the various models produce just analytic 
continuations from values close to the nuclear surface. Note that the three 
extrapolated imaginary KM$\alpha$ potentials are deeper than the purely 
phenomenological imaginary potential. It is reassuring that the SNAF are 
largely determined by Im$\,V_{K^-}$ contributions from densities 10-20\% of 
$\rho_0$ where the balance between $V_{K^-}^{1N}$ and $V_{K^-}^{mN}$ is 
unambiguously related to the $1N$ model used. 

\begin{figure}[htb] 
\begin{center} 
\includegraphics[width=0.45\textwidth,height=6.1cm]{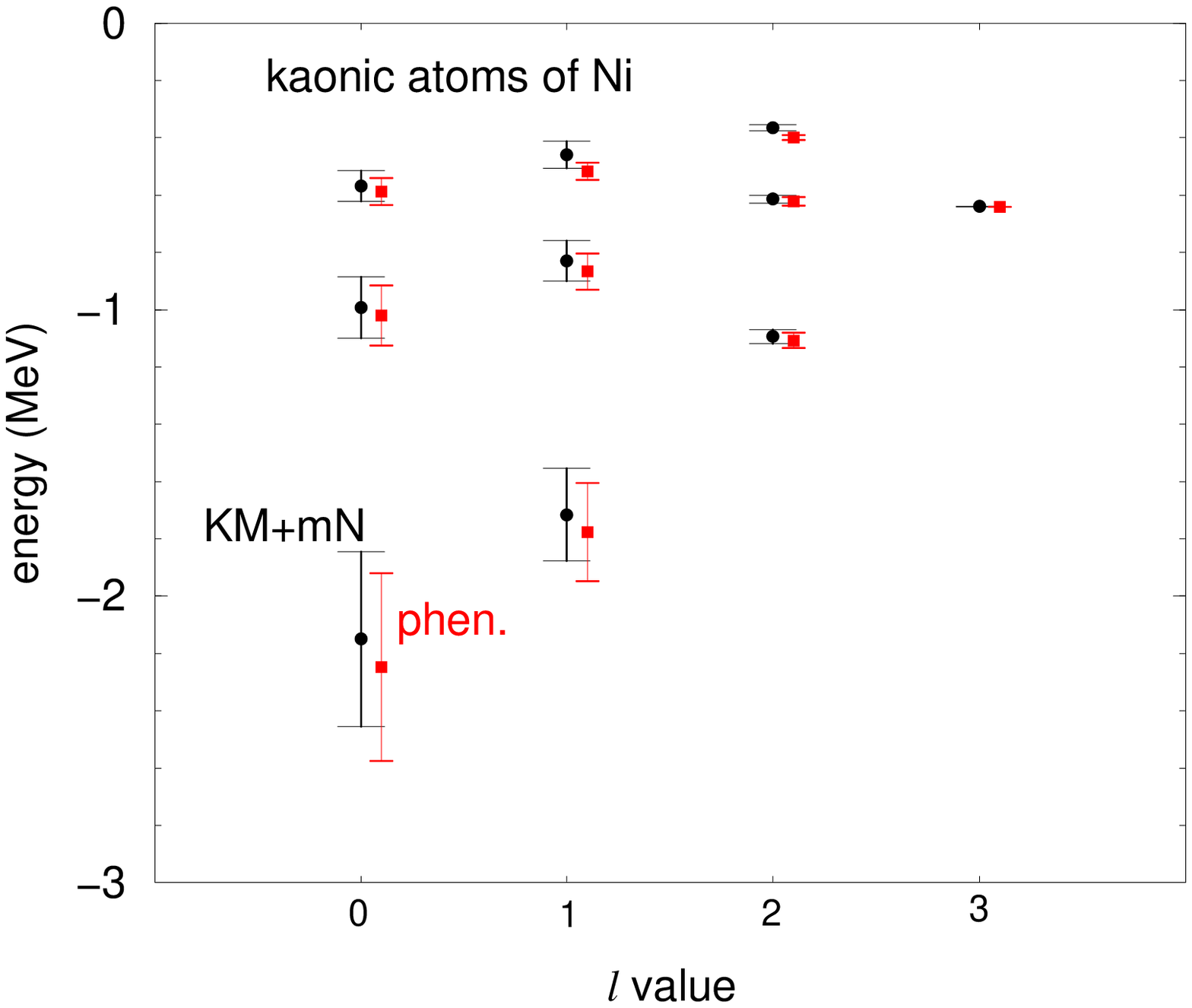} 
\includegraphics[width=0.48\textwidth,height=6.1cm]{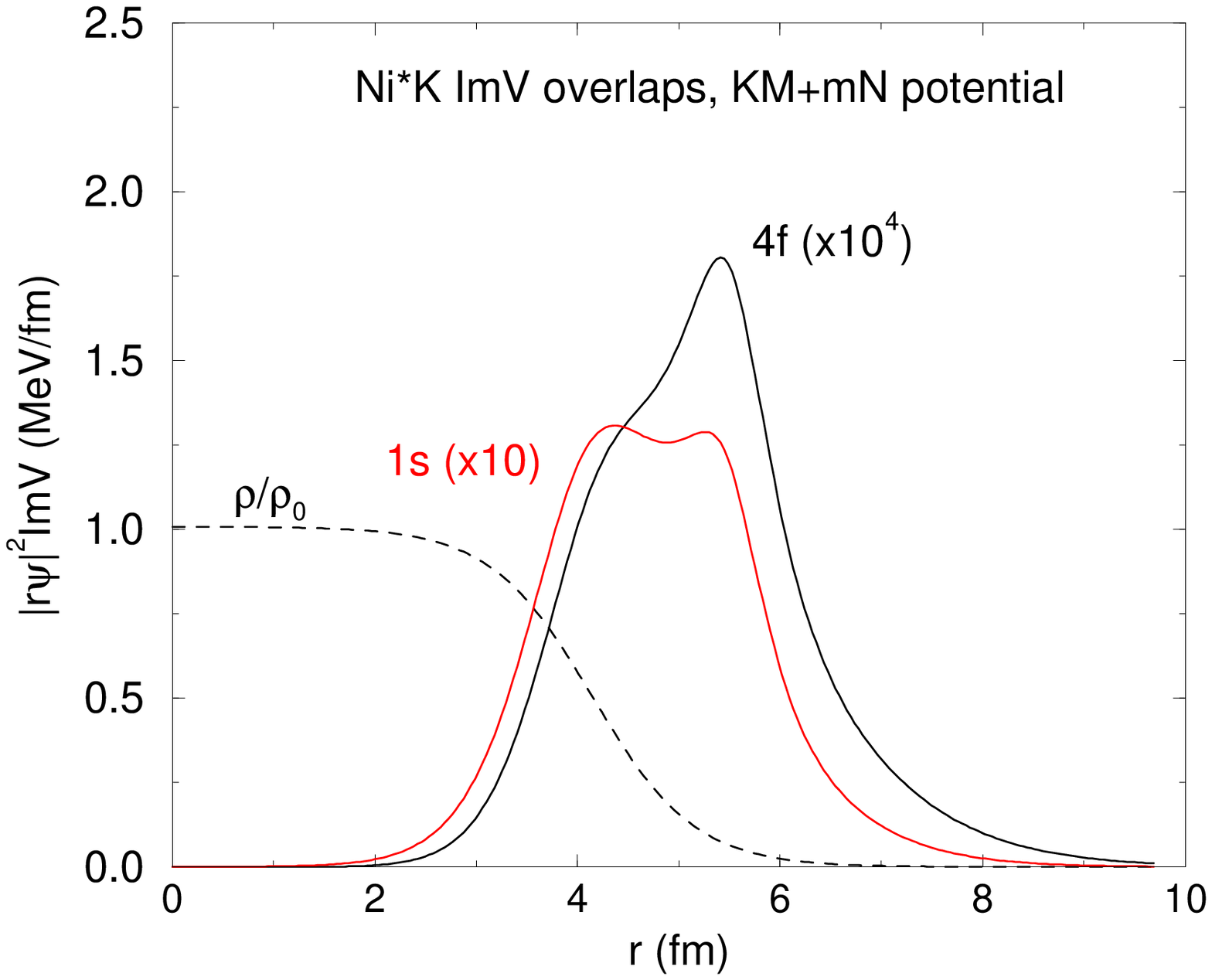} 
\caption{Left: energy levels and widths of kaonic atoms of Ni, 
calculated (in red) for a purely phenomenological best-fit potential 
\cite{FG99a,FG99b} and (in black) for the KM1 optical potential. 
Right: Overlaps of the absolute value squared of the 4f and 1s 
radial wave function in kaonic atoms of Ni with the imaginary part of the 
KM1 optical potential. Shown in dash is the relative nuclear density 
of Ni.} 
\label{fig:deeply} 
\end{center} 
\end{figure} 

Fig.~\ref{fig:deeply} shows spectra (left panel) and overlaps (right panel) 
from Eq.~(\ref{eq:Im}) for kaonic atoms of Ni. The positions, as well as 
widths of energy levels plotted for two optical potentials (see caption) 
nearly coincide, {\it provided} these potentials fit the entire data on kaonic 
atoms. In particular the width of the lowest 1s state is rather small, less 
than 1~MeV, in spite of the large nuclear surface values of Im$\,V_{K^-}\sim 
-50$~MeV from Fig.~\ref{fig:Vopt}. The resolution of this apparent paradox 
is that this same strongly absorptive Im$\,V_{K^-}$ acts as inner repulsion, 
excluding the $K^-$ meson in its 1s atomic state from penetrating 
the nucleus nearly as much as it does in the X-ray observed 4f state, 
as shown on the right panel. It therefore appears that deeply bound 
kaonic atom states~\cite{FG99a,FG99b}, if ever measured in strong-interaction 
production reactions, will not provide information on the interior different 
from that known from normal X-ray states.

\section{$K^-$ nuclear quasibound states} 
\label{sec:nuclei} 

\begin{figure}[htb]
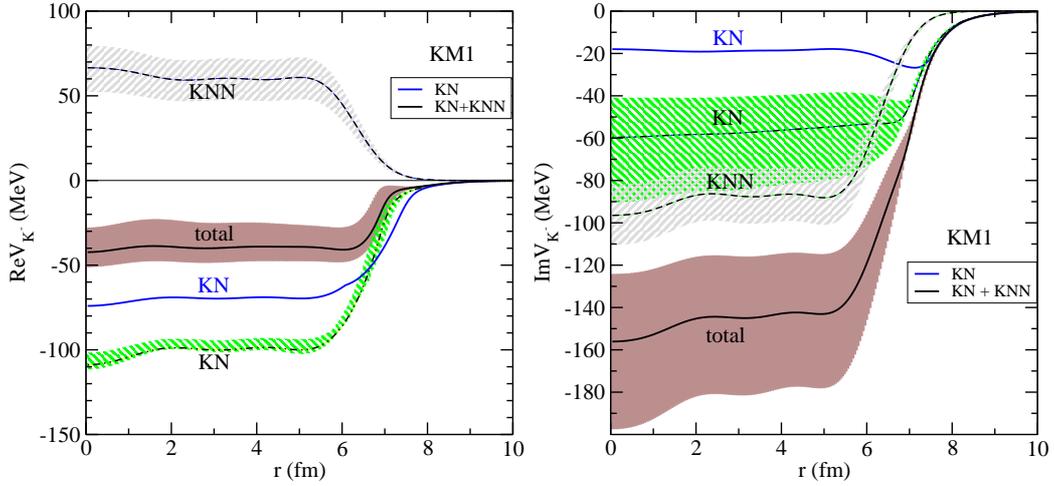
 
\begin{center} 
\includegraphics[width=0.45\textwidth]{ReKM1compare+unc.eps} 
\includegraphics[width=0.45\textwidth]{ImKM1compare+unc.eps} 
\caption{Contributions from $K^-N$ (dash-dotted) and $K^-NN$ (dashed) terms 
to the total real (left) and imaginary (right) $K^-$ optical potential KM1 
for strongly bound $K^-$ states calculated self consistently in $^{208}$Pb. 
Shaded areas denote uncertainty bands. Switching off $V_{K^-}^{mN}$, 
the purely-KM $V_{K^-}^{1N}$ optical potential is shown for comparison 
($KN$, blue solid lines). Figure adapted from Ref.~\cite{HM17b}.} 
\label{fig:KM1KM2Pb} 
\end{center} 
\end{figure} 

In contrast to $K^-$ atomic states, wavefunctions of $K^-$ nuclear states 
are fully confined within the nucleus. The dominance of the $mN$ component 
of $V_{K^-}$, particularly its imaginary part as noted for kaonic atoms in 
Sect.~\ref{sec:atoms} is also realized in most of the models considered in 
Refs.~\cite{HM17a,HM17b} discussing $K^-$ nuclear quasibound states. This is 
demonstrated for the KM1 optical potential in Fig.~\ref{fig:KM1KM2Pb}. Recall 
that of the six model amplitudes of Fig.~\ref{fig:Ffree}, P and KM are the 
only ones passing the test of producing realistic values of SNAF~\cite{FG17}. 
The $1N$ (denoted $KN$) and $mN$ (denoted $KNN$) potential contributions 
calculated self consistently in $^{208}$Pb, including their uncertainties, 
are plotted in this figure as a function of the radial distance from the 
nuclear center. The $1N$ component of $V_{K^-}$ is seen to differ from the 
input $1N$ term (blue solid lines) owing to the different subthreshold energy 
shifts obtained upon in(ex)cluding the $mN$ phenomenological term. The marked 
uncertainties reflect the kaonic-atom fit uncertainties in the values of the 
$mN$ strength parameter $B$. Whereas the resulting real potential depths are 
lower than obtained in totally phenomenological analyses~\cite{MFG06}, the 
imaginary potentials that are dominated by the $mN$ component are extremely 
deep, close to 160 MeV (and much higher in the KM2 optical potential). 
Consequently, $K^-$ nuclear quasibound states calculated in these KM-based, 
as well as P-based optical potentials are also extremely broad, with widths 
exceeding 100 MeV each. This conclusion holds across the periodic table, 
from $^6$Li to $^{208}$Pb~\cite{HM17b}, leaving room for observation of 
$K^-$ nuclear quasibound states only in the lightest systems such as $K^-pp$.

\section{Conclusion} 

Six chirally-inspired $K^-N$ interaction models~\cite{CMMS16} were used 
in global fits to kaonic atoms~\cite{FG17}. As expected, they all require 
additional purely phenomenological multi-nucleon terms to provide 
state-of-the-art fits. 
Although appearing equivalent on this basis, only two models (P and KM) pass 
the test of reproducing the SNAF extracted in bubble-chamber experiments. 
It was found that the derived $K^-$ optical potentials are meaningful only 
for densities up to 25\% of $\rho_0$ for the real part and up to 50\% of 
$\rho_0$ for the imaginary part. Revisiting deeply bound kaonic 
atoms~\cite{FG99a,FG99b} it was found that such states are well defined, 
but are hardly sensitive to details of best-fit optical potentials. This is 
caused by the poor overlap of $K^-$ atom wave functions with the nucleus, 
which limits the kaon to relatively small nuclear densities and also makes 
these states remarkably narrow. 

For $K^-$ nuclear quasibound states~\cite{HM17a,HM17b} in contrast, the 
overlap of $K^-$ nuclear wavefunctions with the nucleus is substantial. 
Hence, the widths calculated for these states exceed 100~MeV when the 
optical potential $V_{K^-}$ is extrapolated beyond its range of applicability 
of about 0.5$\,\rho_0$. We conclude that a search for such states in nuclei 
heavier than $^4$He appears hopeless.

\acknowledgments 

This work was partly supported by the GACR Grant No. P203/15/04301S.


\begin{thebibliography}{99}

\bibitem{MNP17} V. Metag, M. Nanova, E.Ya. Paryev, Prog. Part. Nucl. Phys. 
{\bf 97} (2017) 199. 

\bibitem{BFG97} C.J. Batty, E. Friedman, A. Gal, Phys. Rep. {\bf 287} (1997) 
385. 

\bibitem{FG07} E. Friedman, A. Gal, Phys. Rep. {\bf 452} (2007) 89. 

\bibitem{CMMS16} A. Ciepl\'{y}, M. Mai, U.-G. Mei{\ss}ner, J. Smejkal, 
Nucl. Phys. A {\bf 954} (2016) 17.  

\bibitem{CFGGM11a} A. Ciepl\'{y}, E. Friedman, A. Gal, D. Gazda, J. Mare\v{s}, 
Phys. Lett. B {\bf 702} (2011) 402. 

\bibitem{CFGGM11b} A. Ciepl\'{y}, E. Friedman, A. Gal, D. Gazda, J. Mare\v{s}, 
Phys. Rev. C {\bf 84} (2011) 045206. 

\bibitem{FG17} E. Friedman, A. Gal, Nucl. Phys. A {\bf 959} (2017) 66. 

\bibitem{HM17a} J. Hrt\'{a}nkov\'{a}, J. Mare\v{s}, Phys. Lett. B {\bf 770} 
(2017) 342. 

\bibitem{HM17b} J. Hrt\'{a}nkov\'{a}, J. Mare\v{s}, Phys. Rev. C {\bf 96} 
(2017) 015205. 

\bibitem{WRW97} T. Waas, M. Rho, W. Weise, Nucl. Phys. A {\bf 617} (1997) 449. 

\bibitem{FG14} E. Friedman, A. Gal, Nucl. Phys. A {\bf 928} (2014) 128. 

\bibitem{MFG06} J. Mare\v{s}, E. Friedman, A. Gal, Nucl. Phys. A {\bf 770} 
(2006) 84. 

\bibitem{FG99a} E. Friedman, A. Gal, Phys. Lett. B {\bf 459} (1999) 43. 

\bibitem{FG99b} E. Friedman, A. Gal, Nucl. Phys. A {\bf 658} (1999) 345. 


\end{thebibliography}
\end{document}